\documentclass[letters,useAMS,usenatbib]{mnras}

\usepackage{amsmath, amssymb}
\usepackage{epsfig}         
\usepackage{graphicx, float}
\usepackage{multirow}
\usepackage{tabularx}
\usepackage{bm}            

\usepackage{natbib}
\usepackage{aas_macros}

\usepackage[]{hyperref}
\usepackage[hyphenbreaks]{breakurl}

\newcolumntype{L}[1]{>{\raggedright\arraybackslash}p{#1}}
\newcolumntype{C}[1]{>{\centering\arraybackslash}p{#1}}
\newcolumntype{R}[1]{>{\raggedleft\arraybackslash}p{#1}}
\usepackage{xcolor,colortbl}
\definecolor{Gray}{gray}{0.85}
\newcolumntype{G}{>{\columncolor{Gray}}r}

\def\kpc{~\rm kpc}

\def\Msun{\rm{M}_{\odot}}

\def\kms{~\rm{km/s}}

\newcommand{\coco}{\textsc{COCO}}
\newcommand{\apostle}{\textsc{APOSTLE}}

\newcommand {\fE} {f_{\rm{E;\; rad}}}

\newcommand {\lcdm}{$\Lambda$CDM}

\newcommand{\eq}[1]{Eq. \eqref{#1}}

\newcommand{\reffig}[1]{Figure \ref{#1}}

\newcommand{\reffigS}[2]{Figures \ref{#1} and \ref{#2}}

\newcommand{\figDir}{fig_pdf/}

\usepackage{color}
\definecolor{mycolor}{rgb}{.0,.3,1.}

\newcommand{\MCn}[1]{#1} 
\newcommand{\MCd}[1]{} 
\newcommand{\MCc}[1]{} 
\newcommand{\MCq}[1]{} 

\title{The tangential velocity excess of the Milky Way satellites}

\author[Cautun $\&$ Frenk]
{\parbox{\textwidth}{
        Marius Cautun$^{1}$\thanks{E-mail : m.c.cautun@durham.ac.uk} 
        and Carlos S. Frenk$^{1}$
        \vspace{.1cm}} \\
$^1$   Department of Physics, Institute for Computational Cosmology, Durham University, South Road Durham DH1 3LE, UK
}

\begin{document}


\maketitle

\begin{abstract}
  We estimate the systemic orbital kinematics of the Milky Way
  classical satellites and compare them with predictions from the
  $\Lambda$ cold dark matter (\lcdm{}) model derived from a
  semi-analytical galaxy formation model applied to high resolution
  cosmological N-body simulations. We find that the Galactic satellite
  system is atypical of \lcdm{} systems.  The subset of 10 Galactic
  satellites with proper motion measurements has a velocity
  anisotropy, $\beta=-2.2\pm0.4$, that lies in the $2.9\%$ tail of the
  \lcdm{} distribution. Individually, the Milky Way satellites have
  radial velocities that are lower than expected for their proper
  motions, with 9 out of the 10 having at most $20\%$ of their orbital
  kinetic energy invested in radial motion. Such extreme values are
  expected in only $1.5\%$ of \lcdm{} satellites systems. \MCn{In the standard cosmological model,} this
  tangential motion excess is unrelated to the existence of a Galactic
  ``disc of satellites". We present theoretical predictions for larger
  satellite samples that may become available as more proper motion
  measurements are obtained.
\end{abstract}

\begin{keywords}
{Galaxy: halo - kinematics and dynamics - Local Group - cosmology: theory - dark matter}
\end{keywords}

\section{Introduction}
\label{sec:introduction}
Several predictions of the current cosmological paradigm -- the 
$\Lambda$ cold dark matter  (\lcdm{}) model -- agree with observations 
such as those of the temperature anisotropies of the cosmic microwave 
background radiation and galaxy clustering \MCn{(e.g. see \citealt{Frenk2012})}. 
Nonetheless, the model has been claimed to be in disagreement with 
some properties of the Local Group satellites. These claims include the
observations that: there are far fewer dwarf galaxies than there are
dark matter substructures \citep[][a discrepancy misleadingly dubbed
the ``missing satellites'' problem]{Klypin1999,Moore1999}; that the
internal structure of the most massive subhaloes is incompatible with
that of known satellite galaxies \citep[the ``too-big-to-fail"
problem;][]{Boylan-Kolchin2011}; and that a large fraction of
satellites seem to rotate in a thin plane \citep[the ``planes of
satellites" problem;][]{Kroupa2005,Pawlowski2013b,Ibata2013}. The first two
``problems" can be resolved by including realistic galaxy formation
models \citep[e.g.][]{Sawala2016a}, but the latter is more
challenging. Systematic studies of the Milky Way (MW) and M31 planes
of satellites show that such configurations are uncommon, with only
${\sim}10\%$ of \lcdm{} galactic-mass systems having more prominent
planes than those in the Local Group \citep{Cautun2015a,Cautun2015b}. 

In this letter, we compare the kinematics of the Galactic satellites
with the predictions of \lcdm{}. We do so for the subset of 10
satellites that have HST proper motions (the 11 classical ones except
Sextans). Previous studies have focused on two aspects of satellite
kinematics: measuring the clustering of the orbital poles and
reconstructing satellite orbits. The orbital poles are more clustered
than an isotropic distribution, with the clustering being largest for
a subset of eight of the 11 classical satellites
\citep{Pawlowski2013b}. Orbit reconstruction is more challenging since
the outcome is sensitive to both the mass and the radial density
profile of the MW halo \citep[e.g.][]{Lux2010,Barber2014}, both of
which are poorly constrained \cite[][and references
therein]{Wang2015a}. This leads to large uncertainties in the
recovered orbits and, thus, a comparison with theoretical predictions
is not very informative. To overcome such limitations, in this study
we compare the velocity anisotropy parameter and the
fraction of kinetic energy in radial motion between observations and
theory. These two quantities are largely insensitive to the mass of
the halo and to its radial density profile.

\vspace{-.3cm}
\section{Data and simulations}
\label{sec:data}

Our observational sample consists of the 10 bright Galactic satellites
that have HST proper motions. These objects and the sources of their
proper motion measurements are: Sagittarius -- \citet{Pryor2010}; LMC
and SMC -- \citet{Kallivayalil2013}; Draco -- \citet{Pryor2015}, Ursa
Minor -- \citet{Piatek2005}, Sculptor -- \citet{Piatek2006}, Carina --
\citet{Piatek2003}, Fornax -- \citet{Piatek2007}, Leo II --
\citet{Piatek2016}; and Leo I -- \citet{Sohn2013}.  We used satellite
distances and heliocentric velocity values from the
\citet{McConnachie2012} compilation. To obtain the radial and
tangential velocity components with respect to the Galactic Centre we
followed the procedure described in \citet{Cautun2015b}. We generate
1000 Monte Carlo realizations of the MW system in which we sample the
satellite positions and proper motions from Gaussian distributions
centred on the most likely values of each quantity and with dispersion
equal to the uncertainties. These are transformed from heliocentric
to Galactic coordinates, with the Monte Carlo realizations used
to compute confidence intervals. The largest uncertainty is in the
tangential velocities, with $1\sigma$ errors varying from $20$ to
$55\kms$ (median value $40\kms$).

The theoretical model is based on the semi-analytic galaxy formation
model of \citet{Henriques2015} applied to the Millennium II \lcdm{}
dark matter cosmological simulation \citep{Boylan-Kolchin2009}, which
has been rescaled to correspond to the Planck-1 values of the
cosmological parameters \citep[for details see][]{Henriques2015}. Our
sample consists of haloes in the mass range, $M_{200}=(0.8 -
3.0)\times 10^{12}\Msun$, where $M_{200}$ is the mass enclosed by a
spherical overdensity of 200 times the critical density. Our results are 
insensitive to the host halo mass, so we use a broad
mass range motivated by the large uncertainties in the MW halo mass
\citep[][]{Wang2015a} and the advantages of having a large sample of
MW analogues. We find 3672 such host haloes. We restrict the satellite
selection to galaxies with a minimum stellar mass of $10^5\Msun$ found
within a distance of $300\kpc$ from the central galaxy. For each host,
we select the 10 satellites with the largest stellar mass. In the case
of the MW observations, we have proper motions for 10 satellites out
of 12 objects brighter than $M_\rmn{V}=-8.6$ (the classical satellites
and Canes Venitici). To check for systematic biases, we constructed a
second satellite catalogue by randomly selecting 10 out of the 12
objects with the largest stellar mass. We found that the two
catalogues have the same satellite velocity distribution, so, for
simplicity, we limit our analysis to the 10 brightest satellites.

We construct mock satellite catalogues to account for the
uncertainties in the radial and tangential velocity components. We
start by ranking the satellites according to their distance from the
central galaxy. We do the same for the MW satellites. Then, the
simulated satellites are assigned the errors corresponding to the MW
satellite with the same rank, e.g. the innermost satellite in the
simulation is linked to the MW innermost one. To model observational
uncertainties, for every satellite we add to each velocity component a
random value generated from a Gaussian distribution centred on zero
with dispersion equal to the error reported for that velocity
component. We repeat this procedure 10 times for each host, resulting
in 36720 MW mocks.

\vspace{-.3cm}
\section{Results}
\label{sec:results}
\begin{figure}
     \centering
     \includegraphics[width=1.\linewidth,angle=0]{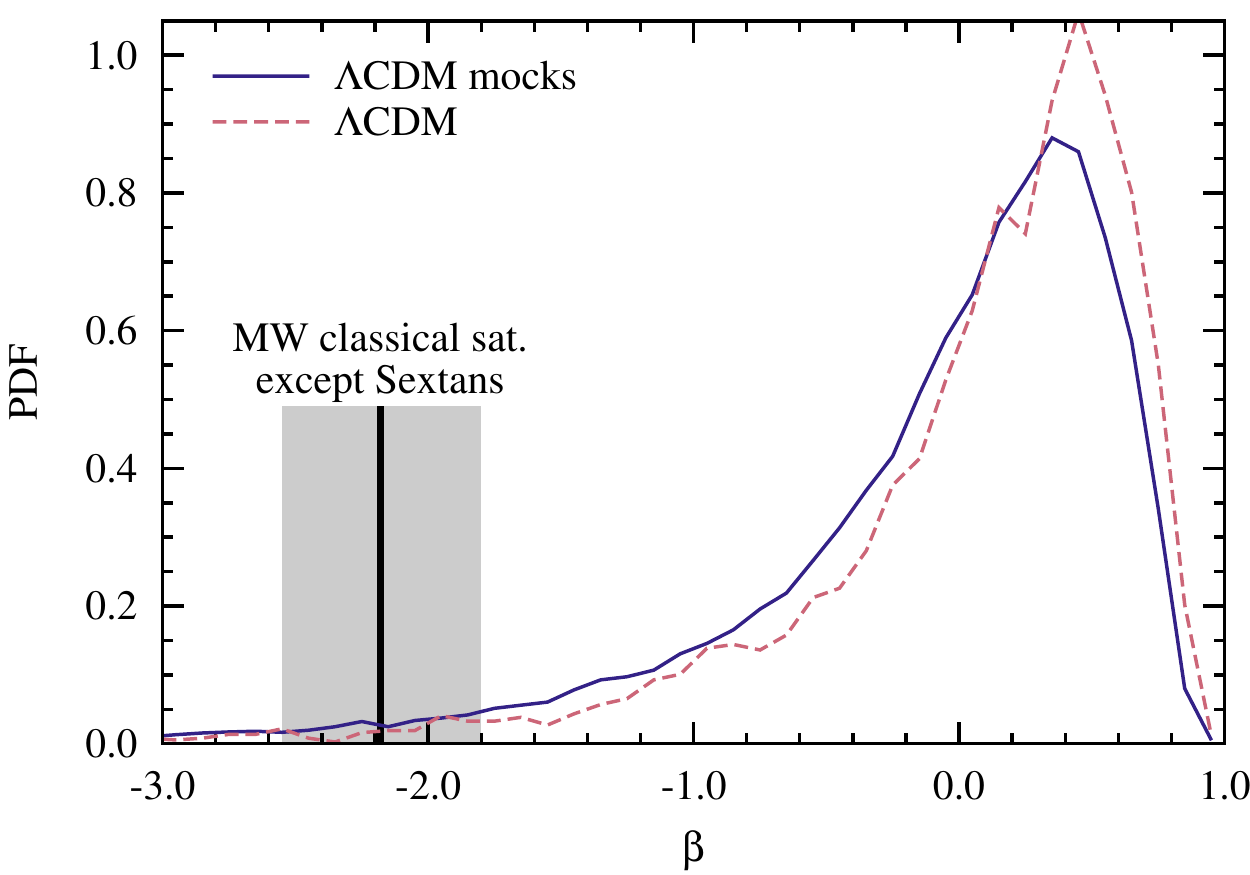}
     \vspace{-0.5cm}
     \caption{ The distribution of the velocity anisotropy, $\beta$,
       for the 10 brightest satellites of MW-mass haloes. We show
       results for the cosmological simulation (dashed line) and for
       mock satellite catalogues that account for observational
       uncertainties (solid line). The vertical line 
       shows the measured value, $\beta=-2.2\pm0.4$, for the MW
       satellites and the grey shaded region shows the $1\sigma$ uncertainty interval. 
       Only $2.9\%$ of mock systems have a lower value of
       $\beta$ than the MW system. }
     \label{fig:velocity_anisotropy}
\end{figure}

The velocity anisotropy parameter, $\beta$, provides a simple measure
of the kinematical properties of satellite galaxies. It is defined as: 
\begin{equation}
    \beta = 1 - \frac{\sum_i V_{\rm tan;\; i}^2}{2\sum_i V_{\rm rad;\; i}^2}
    \label{eq:velocity_anisotropy} \;,
\end{equation}
where $V_{\rm rad;\; i}$ and $V_{\rm tan;\; i}$ denote the radial and
tangential velocity components of satellite $i$ with respect to the
central galaxy. The sum is over all the satellites associated with a
host halo, which, in our case, is 10. The $\beta$ parameter takes values
in the range $-\infty$ to 1, with $\beta < 0$, $\beta=0$ and $\beta>0$
describing circularly-biased, isotropic and radially-biased orbits,
respectively.

\reffig{fig:velocity_anisotropy} shows the distribution of $\beta$
values for the 10 brightest satellites of galactic mass haloes in our
sample. We show the distribution for mock satellite catalogues and
also for the original cosmological simulation (i.e. in the absence of
velocity errors). In both cases, the satellite systems have
radially-biased orbits, with a most likely value, $\beta\simeq0.4$,
but the $\beta$ distribution in the mock catalogues is slightly
shifted towards lower values. The shift is due to the transverse
velocity errors being an order of magnitude larger than the radial
velocity errors. On average, this leads to an overestimation of
$V_{\rm tan;\; i}^2$ by a larger amount than of $V_{\rm rad;\; i}^2$,
and thus a systematic reduction in $\beta$.

The Galactic satellites have $\beta=-2.2\pm0.4$, which means that they have
tangentially-biased motions. \MCn{This agrees with previous studies 
that, using fewer Galactic satellites with HST proper motions,
also found a preference for tangential motions \citep[e.g.][]{Watkins2010,Pawlowski2013b}.}
The $\beta$ value of the Galactic satellites, marked with  a 
vertical line in \reffig{fig:velocity_anisotropy}, lies in the tail of
the theoretical prediction, with only $2.9\%$ of \lcdm{} mock
catalogues having an even more extreme value.

\begin{figure}
     \centering
     \includegraphics[width=1.\linewidth,angle=0]{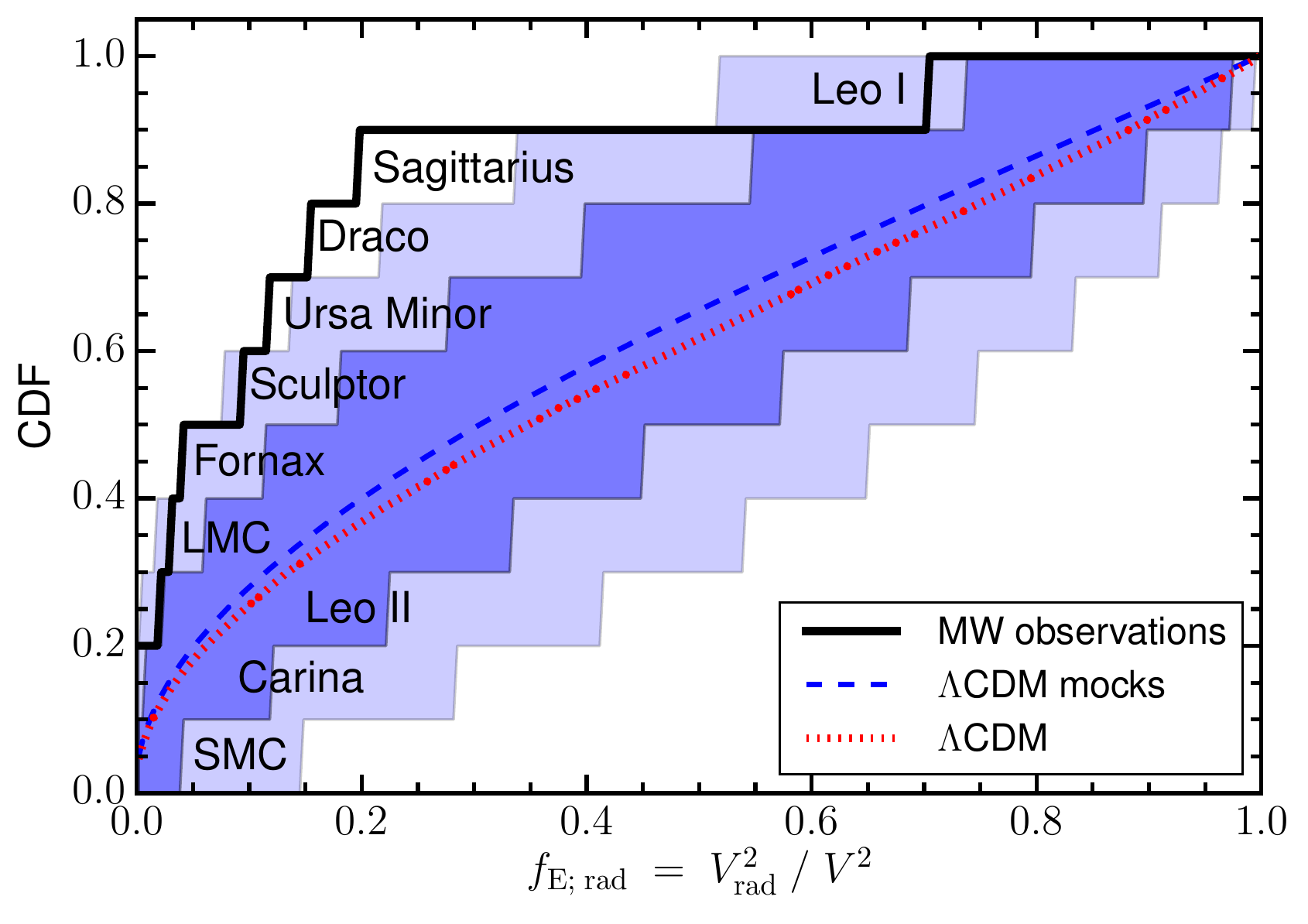}
     \vspace{-0.7cm}
     \caption{ The distribution of the kinetic energy fraction in
       radial motion, $\fE=\tfrac{V^2_\rmn{rad}}{V^2}$, for the 10
       brightest  satellites. The dashed line shows the median trend
       for the \lcdm{} Galactic mocks. The darker and lighter shaded
       regions show the $1-$ and $2$-$\sigma$ scatter regions. The
       distribution of MW satellites, which is shown by the solid
       line, is consistent with the mocks at the $1.5\%$ level. We
       also show the median expectation in the absence of
       observational errors 
       (dotted line). } 
     \label{fig:velocity_ratio}
\end{figure}

\reffig{fig:velocity_ratio} shows the distribution of tangential
versus radial motion for individual satellites. We characterize this
by the fraction of kinetic energy, $\fE=\tfrac{V^2_\rmn{rad}}{V^2}$,
along the radial direction. A satellite that, at a given moment, has a
preferentially tangential motion corresponds to $\fE<\tfrac{1}{3}$,
while a satellite that has a preferentially radial motion corresponds
to $\fE>\tfrac{1}{3}$.  \lcdm{} predicts that at any moment $49\%$ of
the satellites have $\fE<\tfrac{1}{3}$, which increases to $52\%$ for
the Galactic mock satellite
catalogues. 

The distribution of $\fE$ values for the Galactic satellites is
dominated by tangential motions, with $\fE<0.2$ for 9 out of the 10
satellites (thick solid line in \reffig{fig:velocity_ratio}); on
average \lcdm{} predicts only 4 such objects. To quantify the
significance of the disagreement between observations and theory we
cannot just compute the fraction of mock catalogues that have 9 or
more satellites with $\fE=0.2$ since this would be an \textit{a
  posteriori} defined test that disregards the \textit{look elsewhere}
effect. This problem can be overcome by performing an extended
Kolmogorov-Smirnov test that accounts for additional sources of
scatter beyond just those due to Poisson statistics. We define the
maximum difference between the cumulative distribution functions (CDF)
of the data, CDF$_{\rm data}$, and of the mean for the mock catalogues,
CDF$_{\rm mean}$, as: 
\begin{equation}
    D = \max_{\fE} \left| \rm{CDF}_{\rm data}(\fE) -  \rm{CDF}_{\rm mean}(\fE) \right |
    \label{eq:supremum_difference} \;.
\end{equation} 
The Galactic satellite system has $D_{\rm MW}= 0.5$. For each mock
catalogue, we compute $D$ given by \eq{eq:supremum_difference} with the
CDF of the data replaced by the CDF of the $\fE$ values in that particular mock
catalogue. The probability of obtaining a deviation as extreme as that
observed in the data is given by the fraction of mock catalogues with
$D$ values larger than $D_{\rm MW}$. Only $1.5\%$ of mock catalogues
show a larger deviation than the data.

\vspace{-0.3cm}
\section{Discussion}
\label{sec:discussion}

\begin{figure}
     \centering
     \includegraphics[width=1.\linewidth,angle=0]{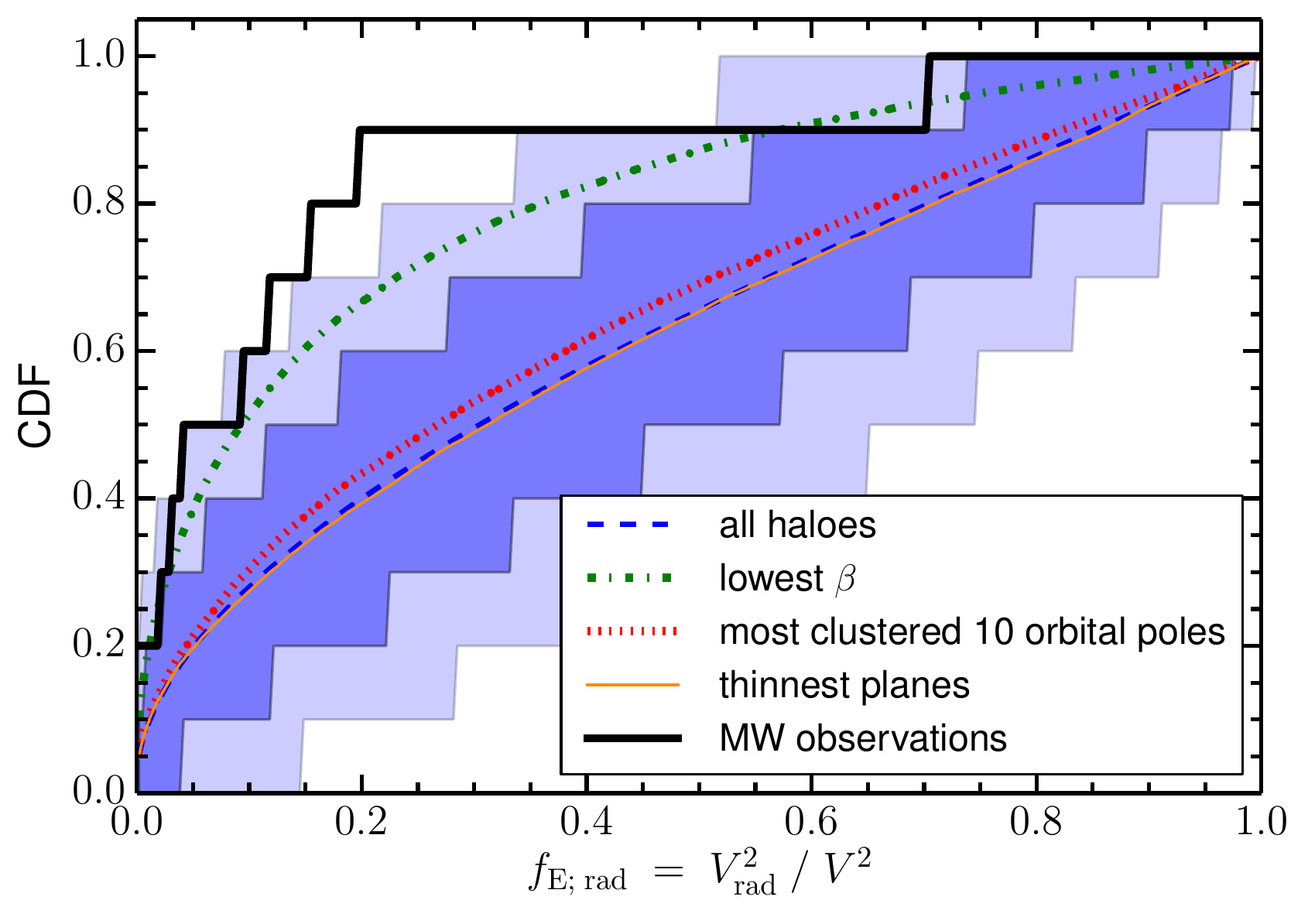}
     \vspace{-0.7cm}
     \caption{ As \reffig{fig:velocity_ratio}, but showing the
       median expectation for all \lcdm{} haloes (dashed line) and for
       subsets consisting of the $5\%$ of haloes that have: the lowest
       $\beta$ values (dashed-dotted line), the
       most clustered satellite orbital poles (dotted line) and the thinnest
       satellite planes (solid thin line). The plot shows that low
       $\beta$ values are highly correlated with low $\fE$ values and
       thus both are expressions of the same phenomenon. The presence
       of a satellite plane or of coherent rotation has little effect
       on the $\fE$ values and thus the two effects are largely
       independent. } 
     \label{fig:velocity_ratio_planes}
\end{figure}

The 10 MW satellites with measured proper motions have tangentially
biased motions to an extent rarely found in \lcdm{}. Only $2.9\%$ of
\lcdm{} systems have lower values of $\beta$ than the MW satellite
system. Even fewer, $1.5\%$, show deviations in the CDF of $\fE$ that
are as extreme as those measured for the MW. The two discrepancies are
expressions of the same property, as may be seen in
\reffig{fig:velocity_ratio_planes}. Selecting the $5\%$ of mock
catalogues with the lowest values of $\beta$ results in a distribution
that is biased towards low $\fE$ values, similarly to that measured in
the real data. In the following, we consider possible reasons behind
the disparity between observations and theory. We focus on the test
illustrated in \reffig{fig:velocity_ratio}, i.e. the CDF of $\fE$,
since that test shows the largest discrepancy and thus is the most
constraining.

In \reffig{fig:velocity_ratio_planes} we investigate whether the
preference for tangential motions is somehow related to the Galactic
``disc of satellites" \citep{Lynden-Bell1976,Kroupa2005,Libeskind2005}. 
Very few haloes have satellite systems similar to that in the MW
\citep{Pawlowski2014c}, i.e. that are as thin and have highly
clustered orbital poles, so, to have good statistics, we need to study
each of these two aspects separately. First we select the $5\%$ of
mock satellite catalogues that have the thinnest planes of
satellites. These are the systems with the smallest values of $c/a$,
where $a$ and $c$ are respectively the major and minor axes of the
inertia tensor of the satellite distribution. This subsample of
haloes, shown with a thin solid line in
\reffig{fig:velocity_ratio_planes}, has the same CDF of $\fE$ values
as the overall sample. Thus, the flattening of a
satellite distribution is uncorrelated with its degree of tangential
motion.

Of the 10 Galactic satellites with measured proper motions, 7 have
orbital poles that are significantly clustered on the sky
\citep{Pawlowski2013b}. For each of our mock satellite systems we
identify the set of $n$ satellites (out of 10) that have the most
strongly clustered orbital poles, i.e. the smallest angular dispersion
in the direction of the orbital poles \citep[see Eq. 6
in][]{Cautun2015b}. For each value of $n$, we then select the $5\%$ of
haloes that have the most clustered orbital poles. These subsamples
show a small preference for tangential motions compared to the full
sample, with the excess being largest for $n=10$. The dotted line in
\reffig{fig:velocity_ratio_planes} shows the most extreme case,
$n=10$. The MW data are consistent with this subsample at the $3.3\%$
level.  The weak correlation between preferentially tangential motions
and clustering of the orbital poles may not be very relevant for the
Galactic satellites. When considering the clustering of all satellites
(i.e. $n=10$), the MW is in the 15th, not the 5th, percentile of the
distribution; mock catalogues corresponding to that percentile behave
exactly like the full sample. The Galactic satellites are extreme for
$n=7$, but the shift in the CDF in that case is much smaller than for
the $n=10$ case shown in \reffig{fig:velocity_ratio_planes}. Thus, a
strong clustering of the orbital poles of satellites is, at
most, only weakly associated with an excess of tangential
motion.

The LMC and the SMC are thought to have been accreted as a pair
\citep[e.g.][]{Besla2012}, so the two galaxies could have correlated
orbital dynamics. This is unlikely to explain the tangential velocity excess
of the Galactic satellites since group
accretion is common in \lcdm{}: when studying the 11 brightest 
satellites, \citet{Wang2013} found accretion of satellite groups with
2 or more members in 5 out of their 6 haloes. Nonetheless, we tested
for the effect of group accretion by excluding the SMC from the
sample. We repeated the analysis for systems of 9 satellites
and found only a small reduction in the difference between data and
theory: the 9 MW satellites lie in the $3.0\%$ tail of the
distribution for \lcdm{}. 
\begin{figure}
     \centering
     \includegraphics[width=.95\linewidth,angle=0]{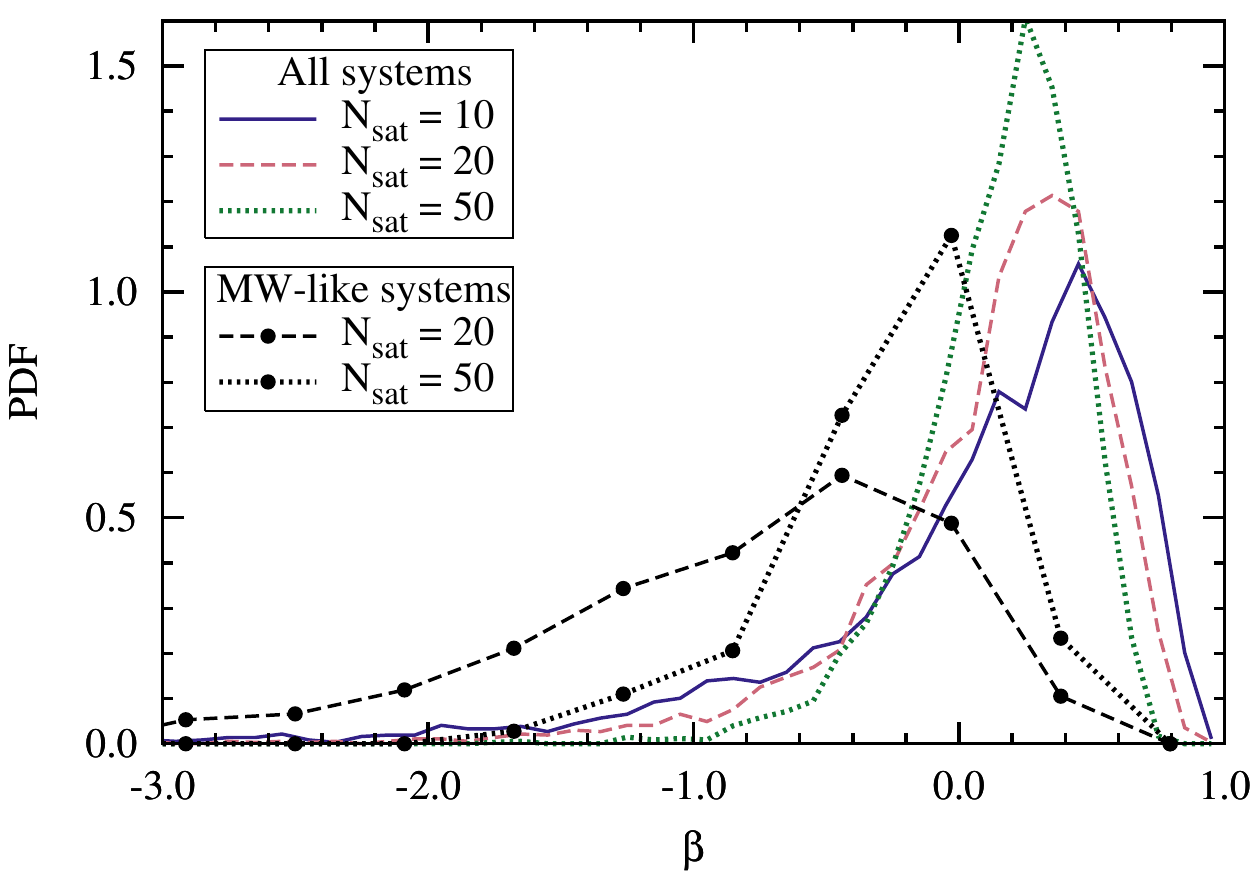}
     \vspace{-0.3cm}
     \caption{ The distribution of velocity anisotropy, $\beta$, for
       the 10 (solid line), 20 (dashed line) and 50 (dotted
       line) brightest satellites of \lcdm{} Galactic-mass haloes. The lines with
       symbols show the same distribution but for haloes selected to
       resemble the MW, that is the $5\%$ of haloes whose 10 brightest 
       satellites have the lowest values of $\beta$. The results do
       not include velocity errors. }
     \label{fig:future_beta}
     \vspace{-0.3cm}
\end{figure}

\begin{figure}
     \centering
     \includegraphics[width=.935\linewidth,angle=0]{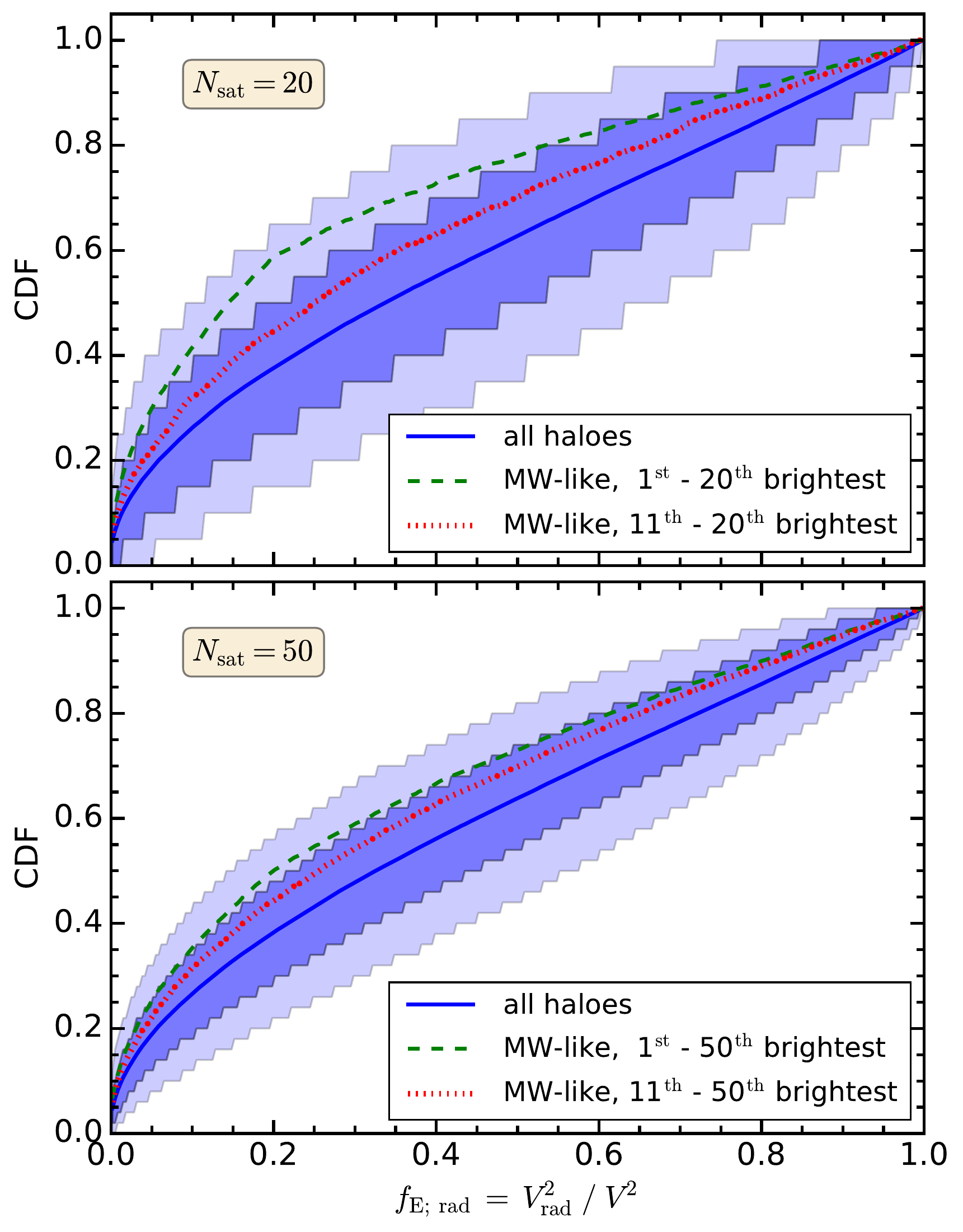}
     \vspace{-0.3cm}
     \caption{ The distribution of the fraction of kinetic energy in
       radial motions, $\fE$, for the 20 (top panel) and
       50 (bottom panel) brightest satellites of \lcdm{} Galactic-mass
       haloes. The darker and lighter shaded regions show the $1-$ and
       $2\sigma$ scatter regions. The dashed line shows the
       expectation for haloes selected to resemble the MW, that is
       haloes for which at least 7 of the 10 brightest satellites have
       $\fE \leq 0.2$ (which corresponds to ${\sim}5\%$ of the
       population). 
       The dotted line shows the expectation for the MW-like systems
       when excluding the 10 brightest satellites, i.e. when
       considering only the $11^{\rm th}-20^{\rm th}$ or the $11^{\rm
         th}-50^{\rm th}$ brightest samples. The results do not
       include velocity errors. }
     \label{fig:future_velocity_ratio}
     \vspace{-0.3cm}
\end{figure}

Satellite proper motions are difficult to measure and could
potentially be affected by unknown systematic errors. \MCn{To reduce the Galactic tangential velocity excess
to a $1\sigma$ disagreement,} the proper
motion of each satellite would have to be overestimated by $45\%$. 
\MCn{Recently, \citet{Casetti-Dinescu2016} published a new ground-based proper motion measurement for Draco that is ${\sim}6\sigma$ discrepant from the Pryor et al. HST measurement. The ground-based measurement gives a much lower tangential velocity, $(90\pm 16)\kms$, compared to the HST value, $(210\pm 25)\kms$. Taking this value would ease the discrepancy between theory and observations, with only 8 out of the 10 Galactic satellites having $\fE\leq0.2$, which would make the MW system a $9\%$ outlier. It remains to be determined by future observations which of the two Draco proper motion measurements is correct and whether the HST measurements are affected by as yet unknown systematic errors.}
Two other concerns might be the limited
resolution and the absence of baryonic effects in the cosmological
simulation used here. We checked for these possibilities by analysing
the \coco{} simulation \citep[][which has $100$ times better mass
resolution]{Hellwing2016a} and the \apostle{} Local Group simulations
\citep[][which include realistic baryonic physics]{Sawala2016a}. We
found good agreement between the results of these simulations and
those of the one used in this study.

The fraction of kinetic energy invested in radial motions, $\fE$,
depends on the position of the satellite along its orbit, being
smallest at pericentre and apocentre. The low $\fE$ value found for
the Galactic satellites could be interpreted as implying that 9 of the
10 satellites are close to either pericentre or apocentre to a larger
extent than is normally found in \lcdm{}. The $\fE$ value also depends on
the orbital ellipticity, being smaller for circularly biased
obits. Thus, the discrepancy between data and theory could
alternatively indicate that the Galactic satellites have orbits that
are, on average, closer to circular than is typical in \lcdm{}. This
would mean that MW halo mass estimates based on satellite orbits
\citep[e.g.][]{Barber2014} are biased low.

More observations are required to decide which, if any, of the above
explanations is correct, or, alternatively, if the excess of
tangential motions is an indication of new physics in the dark
sector. There are two main directions in which this analysis can be
extended: measuring proper motions for fainter Galactic satellites or
performing similar tests for external
galaxies. \reffigS{fig:future_beta}{fig:future_velocity_ratio} show
theoretical predictions for the expected behaviour of $\beta$ and $\fE$ 
as proper motion measurements become available for a larger
number of satellites. The Gaia mission will reduce the uncertainties
in the proper motions of several of the classical satellites and
should obtain new measurements for fainter objects, especially for
those within ${\sim}100\kpc$ from the Sun \citep{Wilkinson1999}. The
proper motions of more distant Galactic satellites and of those in M31
could be measured by a dedicated multi-year HST programme and by
follow-up with JWST and WFIRST \citep{Kallivayalil2015}.

\reffig{fig:future_beta} shows that the velocity anisotropy, $\beta$,
decreases as fainter satellites are included in the sample, with the
typical value varying from $\beta=0.45$ for the $10$ brightest
satellites to $\beta=0.25$ for the $50$ brightest satellites. The
distribution of $\beta$ becomes more peaked and narrower for larger
satellite samples. To make predictions constrained by the already 
existing data for the MW, where the 10 brightest satellites have 
very low $\beta$ values, we select the $5\%$ of haloes whose 10 
brightest satellites have the lowest velocity anisotropy 
($\beta\leq-1.3$). The velocity anisotropy of the
20 or 50 brightest satellites remains biased low for these systems
relative to the full sample of haloes.

In \reffig{fig:future_velocity_ratio} we show the CDF of the kinetic
energy fraction in radial motion for the 20 and 50 brightest 
satellites. While the median trend hardly changes, the scatter is much
reduced as the number of satellites increases. The dashed curves show
the expected behaviour for \lcdm{} systems chosen to be similar to the
MW, that is in haloes for which at least 7 of the 10 brightest satellites have
$\fE \leq 0.2$. These MW-like systems show systematically larger tangential motions
even when excluding the 10 brightest satellites, which were used in
the first place to select the sample.

\vspace{-0.3cm}
\section{Summary}
\label{sec:summary}

We have found that the bright satellites of the MW have larger
tangential orbital motions than expected from \lcdm{} cosmological
simulations.  This excess is most clearly manifest in the fraction of
kinetic energy along the radial direction, $\fE$, with 9 of the 10 MW
satellites with HST proper motion measurements having $\fE<0.2$. Such
extreme values are found in at most $1.5\%$ of \lcdm{} galactic
satellite systems. 
\MCn{This conclusion, of course, relies on the accuracy of current HST proper motion measurements, which has been called into question by a recent measurement of Draco using ground-based data.}
\MCn{In \lcdm{},} the tangential motion excess is unrelated to the
existence of a Galactic ``disc of satellites" and cannot be explained
by the accretion of satellite groups.
More satellites with
measured proper motions are required to check if the observed excess
is merely an indication that the MW is atypical or if it  poses a
problem for the \lcdm{} model.

\vspace{-0.2cm}
\section*{Acknowledgements}
We thank Andrew Cooper, Alis Deason, Mark Gieles and Till Sawala for helpful discussions and comments.
This work was supported in part by ERC Advanced Investigator grant COSMIWAY 
[grant number GA 267291] and the Science and Technology Facilities Council 
[grant number ST/F001166/1, ST/I00162X/1].
This work used the DiRAC Data Centric system at Durham University, 
operated by ICC on behalf of the STFC DiRAC HPC Facility (www.dirac.ac.uk). 
This equipment was funded by BIS National E-infrastructure capital 
grant ST/K00042X/1, STFC capital grant ST/H008519/1, and STFC DiRAC 
Operations grant ST/K003267/1 and Durham University. DiRAC is part 
of the National E-Infrastructure. Data from the Millennium-II 
simulation is available on a relational database accessible from 
\url{http://gavo.mpa-garching.mpg.de/Millennium/} .

\vspace{-0.4cm}
\bibliographystyle{mnras}
\bibliography{sat_references}

\end{document}